# Morphology and properties evolution upon ring-opening polymerization during extrusion of cyclic butylene terephthalate and graphene-related-materials into thermally conductive nanocomposites

S. Colonna[a], O. Monticelli[b], J. Gomez[c], G. Saracco[d], A. Fina[a],*

[a]Dipartimento di Scienza Applicata e Tecnologia, Politecnico di Torino, 15121 Alessandria, Italy

[b]Dipartimento di Chimica e Chimica Industriale, Università di Genova, 16146 Genova, Italy

[c]AVANZARE Innovacion Tecnologica S.L., 26370 Navarrete (La Rioja), Spain

[d]Istituto Italiano di Tecnologia, Centre for Sustainable Futures CSF@PoliTo, 10129 Torino, Italy

*Corresponding author: alberto.fina@polito.it

**Abstract**

In this work, the study of thermal conductivity before and after *in-situ* ring-opening polymerization of cyclic butylene terephthalate into poly (butylene terephthalate) in presence of graphene-related materials (GRM) is addressed, to gain insight in the modification of nanocomposites morphology upon polymerization. Five types of GRM were used: one type of graphite nanoplatelets, two different grades of reduced graphene oxide (rGO) and the same rGO grades after thermal annealing for 1 hour at 1700°C under vacuum to reduce their defectiveness. Polymerization of CBT into pCBT, morphology and nanoparticle organization were investigated by means of differential scanning calorimetry, electron microscopy and rheology. Electrical and thermal properties were investigated by means of volumetric resistivity and bulk thermal conductivity measurement. In particular, the reduction of nanoflake aspect ratio during ring-opening polymerization was found to have a detrimental effect on both electrical and thermal conductivities in nanocomposites.

1. Introduction

The high corrosion resistance, ease of processing, lightweight and low cost of polymers attracted scientific community and industry for the preparation of thermally conductive materials for the replacement of metallic parts [1-3], despite their intrinsic low thermal conductivity. For this purpose, polymer composites or nanocomposites are typically used, exploiting different types of conductive fillers/nanofillers, including carbon nanotubes, graphite, graphene-related materials, boron nitride, metal particles, aluminum oxide, diamond and silicon nitride [3].

In the last decade, scientists focused their attention on graphene [4, 5], a single-atom-thick sheet of hexagonally arranged $sp^2$-bonded carbons, due to its extraordinary thermal, electrical and mechanical properties [6-8]. Unluckily, although various preparation approaches have been studied [5, 9-12], the industrial scale up of graphene, required for its exploitation in polymer nanocomposites, remains highly challenging. For this reason, different synthesis techniques have been developed during years: chemical reduction of graphene oxide (GO) [13], thermal exfoliation and reduction of GO [14], ball milling [15], liquid phase exfoliation of graphite [11] and other methods [16]. However, all of these methods often lead to flakes of relatively low quality (low to limited single layer yield, large thickness distribution, chemical defectiveness in the $sp^2$ structure, limited lateral size). Indeed, techniques allowing kg scale production typically lead to so-called graphene-related materials (GRM), including graphite nanoplatelets (GNP), reduced graphene oxide (rGO), multilayer graphene (MLG), which sometimes are misreferred to graphene, while a more accurate nomenclature should be used [17].

Despite the obvious scatter of results, owing mainly to their different structural features, graphene-related materials appear to be more efficient than carbon nanotubes (CNT) for the preparation of thermally conducting polymer nanocomposites [18]. However, for carbon based material families, parameters like particle aspect ratio and quality, interfacial thermal resistance, distribution and alignment of nanoparticles are generally recognized to play a key role for the final thermal conductivity of polymer nanocomposite [1, 2]. The quality of nanoparticles exploited is obviously crucial: while the effect of lateral size is well known [2, 19, 20], in our previous work we demonstrated the increase of the intrinsic thermal conductivity of rGO upon annealing at 1700°C in vacuum [21]. This in turn leads to higher thermal conductivities with respect to nanocomposites containing pristine rGO (2-fold and 4-fold increase when pristine or annealed nanoparticles, respectively, were used) [18].

While homogeneous distribution of conductive particles into the polymer matrix is an obvious need, it is worth noting that nanoparticle dispersion, intended as the separation of nanoparticles into isolated individual primary particle, is expected to be detrimental for conductivity, either thermal or electrical. In fact, the existence of a percolating network is essential for electrical conductivity and generally also recognized as required for efficient thermal transfer, despite heat transfer may also include a role of the matrix between conductive particles. However, when using nanoparticles with a high aspect ratio, e.g. carbon nanotubes or graphene, the percolation threshold is very low and a network of nanoparticles in contact with each other is typically obtained in composites containing a few wt.% nanoparticle loading. Despite this fact, improving contacts between nanoparticles by controlled particle segregation is one of the possible strategies for the enhancement of thermal conductivity in nanocomposites [22]. For instance, Eksik et al. [23] added chemically reduced graphene oxide (c-rGO) coated poly (methyl methacrylate) microspheres to epoxy resin (with a final content of c-rGO of 1 wt.%) and obtained a 7-fold increase in the thermal conductivity of pure epoxy resin while the increase was about 3-fold when a traditional epoxy + 1 wt.% c-rGO nanocomposite was prepared, thus indicating that the formation of a well-organized pathway is crucial for the enhancement of thermal conductivity in polymer nanocomposites.

In the last years, scientists focused their attention on reactive extrusion of polymer nanocomposites, taking advantage of the starting low viscosity of oligomers/monomers and its increase during polymerization reaction, which could result in a higher dispersion and distribution of nanoparticles [24-26]. A wide interest was focused to the *in-situ* ring-opening polymerization (ROP) of cyclic butylene terephthalate (CBT) oligomers for the preparation of poly (butylene terephthalate) (pCBT) nanocomposites [27-30]. In particular, Noh et al. [26] and Colonna et al. [18] demonstrated the possibility to distribute and disperse GNP and rGO via ROP of CBT. The high dispersion and distribution degree of these electrically and thermally conductive nanoparticles resulted in nanocomposites with excellent electrical and thermal conductivities [18, 26]. However, it is worth noting that the presence of graphene related materials was proven to decrease polymerization rate [31] and/or decrease molecular weight [32].

Despite pCBT nanocomposites have been widely studied, the evolution of nanoparticle dispersion and thermal/electrical conductivities during ring-opening polymerizations were not previously reported, to the best of our knowledge. Even though CBT is a solid at room temperature, unpolymerized CBT nanocomposites cannot be regarded as materials for real application owing to their low mechanical properties. Nevertheless, it is important to study how nanocomposite properties change upon the ring-opening polymerization, to be able to develop and optimize reactive extrusion processing of pCBT nanocomposites. With this aim, in the present paper we report a comparison between CBT and pCBT nanocomposite properties, focusing our attention on electrical and thermal properties of nanocomposites containing GNP, rGO and thermally annealed rGO.

## 2. Experimental

### 2.1. Materials

Cyclic butylene terephthalate oligomers [CBT100, $M_w = (220)_n$ *g/mol*, *n* = 2-7, melting point= 130 ÷ 160°C] were purchased from IQ-Holding[*] (Germany). Butyltin chloride dihydroxide catalyst (96%, $m_p$ = 150°C, CAS # 13355-96-9) was purchased from Sigma-Aldrich while acetone (99+%) was purchased from Alfa Aesar.

Three types of graphitic nanoparticles were used for this study. GNP (Surface Area = 22 ± 5 m$^2$/g, Raman $I_D/I_G$ ≈ 0.16, Oxygen content ≈ 1.8 at.%[†], $T_{Oxid}$ ≈ 632°C[‡]), RGO (Surface Area = 210 ± 12 m$^2$/g, Raman $I_D/I_G$ ≈ 0.88, Oxygen content ≈ 3.2 at.%[†], $T_{Oxid}$ ≈ 558°C[‡]) were research grades synthetized by AVANZARE (Navarrete, La Rioja, Spain) accordingly with the procedure previously reported [18]. Another thermally reduced graphite oxide grade, EXG98 350R (Surface Area > 300 m$^2$/g, Raman $I_D/I_G$ ≈ 0.80, Oxygen content ≈ 7.0 at.%[†], $T_{Oxid}$ ≈ 471°C[‡]), was purchased from Graphite Kropfmühl (Germany) and here referred to as RGO-2.

---

[*] Distributor of products previously commercialized by Cyclics Europe GmbH
[†] XPS, $O_{1s}$ signal
[‡] Onset TGA plots in air, 10°C/min heating rate, sample mass ≈ 2 mg

Part of the RGO and RGO-2 were annealed, in a closed graphite box, at 1700 °C for 1 hour at 50 Pa in a vacuum oven (Pro.Ba., Italy) with graphite resistors to decrease the defectiveness in the sp$^2$ structure, as previously reported [21]. Annealed material are from now on referred to as RGO_1700 ($I_D/I_G \approx 0.11$, $O_{1s} \approx 0.4$ at.%, $T_{Onset,TGA} \approx 750°C$) and RGO-2_1700 ($I_D/I_G \approx 0.11$, $O_{1s} \approx 1.1$ at.%, $T_{Onset,TGA} \approx 671°C$).

Raman spectroscopy, XPS and Thermogravimetry results above reported were thoroughly discussed previously [18, 21]

### 2.2. Nanocomposite preparation

Nanocomposites were prepared via a 2-step procedure:

1- About 17.1 g of CBT were partially dissolved in 120 ± 10 ml of acetone for about 1 hour under stirring. Then, 0.9 g of different graphene-related materials (GRM), *i.e.* GNP, RGO, RGO-2, RGO_1700 or RGO-2_1700, were added to the solution and the system underwent a manual mixing for about 5 minutes. The obtained mixture was first dried in a chemical hood for 2 hours then in an oven at 80°C for 8 hours under mild vacuum (~10$^1$ mbar) to remove residual acetone and moisture.

2- CBT nanocomposites were prepared by melt mixing the dried and pulverized CBT/GRM mixture into a co-rotating twin-screw micro-extruder (DSM Xplore 15, Netherlands) for 5 minutes at 100 rpm and 190°C. pCBT nanocomposites were prepared by melt mixing the dried and pulverized CBT/GRM mixture for 5 minutes at 100 rpm and 250°C; then, butyltin chloride dihydroxide catalyst (0.5 wt.% respect to the oligomer content) was added and the process carried out for other 10 minutes (keeping screw speed and temperature constant) to complete CBT polymerization.

## 2.3. Characterization

Morphological characterization of graphene-related materials and CBT/pCBT nanocomposites was performed with a high resolution Field Emission Scanning Electron Microscope (FESEM, ZEISS MERLIN 4248). CBT and pCBT nanocomposites were fractured in liquid nitrogen, then coated with a thin layer (~5 nm) of Chromium.

Differential scanning calorimetry (DSC) was performed on a Q20 (TA Instruments, USA) with a heating rate of 10 °C min$^{-1}$ in the temperature range of 25 ÷ 190 °C and 25 ÷ 250 °C for CBT and pCBT nanocomposites, respectively. The used method consisted in: 1) a heating cycle, required to erase the thermal history of the material; 2) a cooling step, to study the crystallization of CBT and pCBT nanocomposites; 3) a second heating step to evaluate the melting temperature of materials.

All the specimens for rheological, electrical and thermal characterization were prepared by compression molding. No significant preferential orientation of nanoflakes was observed from SEM analysis in the specimen cross-section, as expected taking into account the low shear imposed during compression molding and the high viscosity of the nanocomposites.

Rheological properties of CBT/GRM and pCBT/GRM nanocomposites were evaluated on a strain-controlled rheometer (ARES, TA Instruments, USA) with parallel-plate geometry. The test temperature was controlled by a convection oven, equipped with the instrument. Before each measurement, dried nanocomposites were pressed at 190 °C and 250 °C for CBT and pCBT nanocomposites, respectively, into disks with 25mm diameter and 1 mm thickness. Specimens were further dried at 80 °C in vacuum for 8 h before the measurements. Oscillatory frequency sweeps ranging from 0.1 to 100 rad/s with a fixed strain (set between 0.05 % and 0.1 %, depending on the material, in order to perform experiments in the linear region, as determined by strain sweep tests) were performed in air at 190 and 250 °C (for CBT and pCBT, respectively), to investigate viscoelastic properties of nanocomposites. After sample loading, about 5 min equilibrium time was applied prior to each frequency sweep.

Electrical conductivity (volumetric) was measured with a homemade apparatus on disk-shape specimens (1 mm thickness and 25 mm diameter). The apparatus for the measurement is constituted by:

- A tension and direct current regulated power supply (PR18-1.2A of Kenwood, Japan);
- A numeral table multimeter (8845A of Fluke, Everette/USA) furnished with a digital filter to reduce the noise of the measure;
- A palm-sized multimeter (87V of Fluke, Everette/USA);
- Two homemade brass electrodes: a cylinder (18,5 mm diameter, 55 mm height) and a plate (100 mm side, 3 mm thickness); every electrode has a hole for the connection and a wire furnished with a 4mm banana plug.

The measurement system was based on the multimeter method. Power supply is time to time regulated in current or in voltage to have accurate measurement by both the multimeters, limiting the power dissipated on the specimen. The conductivity value was calculated with the following formula:

$$\sigma = \frac{1}{\rho} = \frac{1}{\frac{V}{I} \cdot \frac{S}{l}} \left[\frac{S}{m}\right]$$

(1)

where $S$ and $l$ are the specimen surface and thickness, respectively; $V$ is the voltage and $I$ the electric current, both read by the apparatus.

Isotropic thermal conductivity tests were carried out on a TPS 2500S by Hot Disk AB (Sweden) with a Kapton sensor (radius 3.189 mm) on disk-shaped specimens (prepared by compression molding of dried nanocomposites) with thickness and diameter of about 4 and 15 mm, respectively. Before each measurements, specimens were further stored in a constant climate chamber (Binder KBF 240, Germany) at 23.0 ± 0.1 °C and 50.0 ± 0.1 % R.H. for at least 48 h before tests. The test temperature (23.00 ± 0.01 °C) was controlled by a silicon oil bath (Haake A40, Thermo Scientific Inc., USA) equipped with a temperature controller (Haake AC200, Thermo Scientific Inc., USA).

### 3. Results and discussion

Differential scanning calorimetry was employed to monitor polymerization of CBT into pCBT and the effect of nanoparticles on melting/crystallization behavior of the polymer matrix. Melting signals of CBT/GRM nanocomposites (Figure S1_a) reveal that nanoparticles do not have influence on CBT melting, with the presence of a small exothermic peak at about 80°C and three endothermic peaks located at ~126, ~153 and ~186°C which are typical for CBT [33]. Furthermore, this indicates that graphene related materials are not able to trigger ring opening polymerization of CBT and a catalyst is crucial to promote oligomer polymerization into pCBT; indeed $2^{nd}$ heating curves on pCBT and pCBT/GRM (Figure S1_c) show no traces of melting peaks of CBT, whereas a new endothermic peak at about 226 °C, related to the melting of the poly (butylene terephthalate), appeared for all the nanocomposites. The absence of CBT melting peaks is not sufficient to prove 100% conversion of CBT; however, conversion up to 97% were reported in literature when CBT were polymerized in similar conditions (205°C, 3 min, in presence of the same catalyst used in this work) [34]. The presence of GRM, drastically changes the crystallization behavior of CBT nanocomposites (Figure S1_b): while pure CBT exhibits only one broad exothermic peak (~85°C) during cooling ramp, CBT + GNP crystallizes with two peaks (~80 and ~111°C). On the other hand, all CBT/rGO nanocomposites are characterized by three crystallization peaks (with the three peaks in the range 66 ÷ 71°C, 87 ÷ 100°C and 115 ÷ 124°C) thus indicating splitting of the crystallization process into multiple peaks which is interpreted as the separated crystallization of the different oligomers, accordingly with their melting behavior described above. This suggests the nucleation activity of GRM is exerted preferentially on higher molecular weight fraction in the mixture of CBT. Nucleating effects by GRM were observed as well in pCBT, with a shift in the crystallization peak from ~199°C for pure pCBT to temperatures in the range 201°C to 211°C, depending on the GRM type (Figure S1_d).

The morphologies of CBT and pCBT nanocomposites were investigated by electron microscopy: representative micrographs of CBT and pCBT nanocomposites containing GNP and RGO (representative for all rGO nanocomposites) are reported in Figure 1. In CBT + 5% GNP (Figure 1a) it is possible to observe regions with large aggregates of nanoflakes, in the range of tens of microns, as well as other areas with a low or negligible content of nanoparticles. This is indeed expected for melt blending of aggregated nanoflakes into low molecular weight liquids, as the extremely low viscosity of CBT did not allow to apply sufficiently high shear forces during mixing to obtain optimal dispersion and distribution degree of nanoparticles. CBT/rGO nanocomposites (in Figure 1b is reported as example CBT + RGO) at low magnification exhibit a better distribution of RGO aggregates (with average lateral size of few dozens of micrometers and average thickness ranging from few to ~ 20 μm) which could not be further separated due to the low viscosity of CBT. However, it is worth noting that low viscosity allows CBT to infiltrate the accordion-like structure of RGO, as observable in Figure 1b. While no clear differences were observed for GNP distribution and dispersion in pCBT (Figure 1c) and CBT (Figure 1a) nanocomposites, in pCBT + RGO nanoflake aggregates were strongly reduced in number and size as compared to CBT + RGO, suggesting a dispersion effect obtained during the polymerization of CBT infiltrated into the accordion-like structure. This can be explained by the progressively increasing applied shear during melt mixing, owed to the viscosity increase during polymerization of CBT into pCBT [35], leading to a significant improvement in dispersion and distribution of nanoparticles. Further insight in the mechanisms of infiltration between nanoflakes and their separation during melt mixing were obtained comparing the above-described polymerization in the presence of rGO with the infiltration of pre-polymerized pCBT as well as with the direct melt blending of pCBT with rGO powder, as discussed in details in SI. Electron micrographs on pCBT nanocomposites prepared by the different methods (Figure S2) evidence clear advantage of solvent assisted premixing compared to direct melt blending, which results in poor distribution and dispersion of RGO.

Nanocomposites prepared by CBT polymerization during melt mixing with RGO-2, RGO_1700 and RGO-2_1700 showed very similar morphology compared to the above described RGO counterpart (for more details view supplementary information, Figure S3).

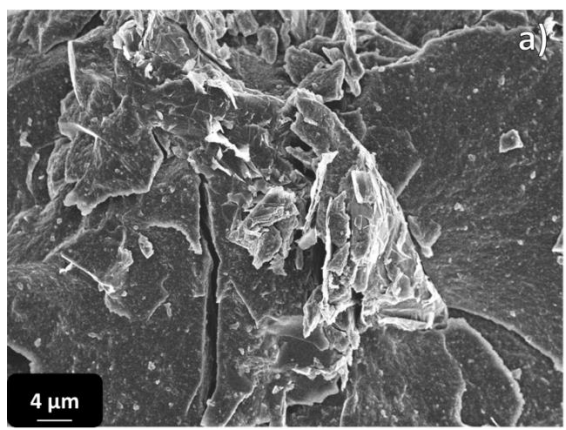
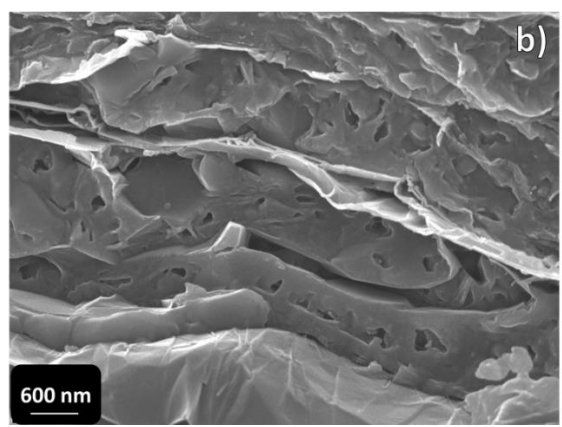

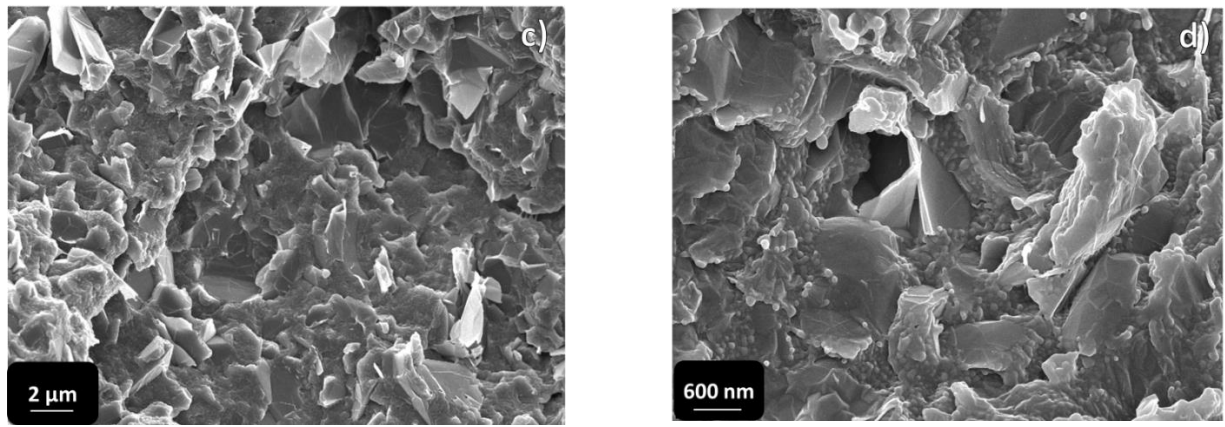

**Figure 1. FESEM micrographs of a) CBT + GNP, b) CBT + RGO, c) pCBT + GNP and d) pCBT + RGO**

Despite electron microscopy is widely used to obtain direct view of composite morphology, it has to be noted that it is clearly a local technique and prone to material homogeneity, nanoparticle orientation and sample preparation method, thus requiring the use of at least another complementary technique to properly assess dispersion and distribution of particles in polymer nanocomposites [36, 37]. For this reason, the study of nanoparticle dispersion and organization in the oligomer and polymer matrixes was completed carrying out dynamic frequency sweep tests in the molten state, as the elastic modulus (G') and complex viscosity ($\eta^*$) of a nanocomposite are strongly affected by nanoflake dispersion and content [38].

$\eta^*$ and G' as a function of deformation frequency plots for pCBT and CBT containing 5 wt.% of GRM are presented in Figure 2: rheological data (G' and $\eta^*$) related to pCBT and its nanocomposites were previously reported [18] and recalled here for comparison with CBT-based nanocomposites. CBT/rGO nanocomposites exhibit a marked dependence of the complex viscosity with frequency [38], with viscosity values several decades higher than those of pure CBT (0.02 Pa·s at 190°C [35, 39]), thus indicating a very strong effect of nanoparticles on the rheology of the oligomers. Furthermore, the weak dependence of G' on the frequency in the whole frequency range used in this work evidences for the formation of a solid-like network of nanoparticles within the molten CBT. It is worth observing that, for a selected type of rGO (RGO or RGO-2), weak differences in η* and G' values (at 190°C) are observed in the whole frequency range when comparing high temperature treated *vs*. untreated particles. In fact, viscosity and G' values (Table S2) of ~ $1·10^5$ Pa·s and ~ $1·10^5$ Pa (at 1 rad/s) were measured for CBT + RGO and CBT + RGO_1700, respectively, while η* and storage modulus values of ~ $7·10^4$ Pa·s and ~ $7·10^4$ Pa (at ω = 1 rad/s) were evaluated for CBT + RGO-2 and CBT + RGO-2_1700, respectively, thus suggesting no effect of nanoflake defectiveness on the rheological properties of their nanocomposites with CBT. On the other hand, the lower viscosity values measured for CBT/RGO-2 with respect to CBT/RGO nanocomposites could suggest a slightly lower dispersion degree when RGO-2 was used. For CBT/GNP nanocomposites, G' and η* values (~ $10^1$ Pa and ~ $10^1$ Pa·s, respectively) as a function of deformation frequency are about 3 to 4 orders of magnitude lower than those measured for CBT + rGO. Despite a significant scattering of data points was observed for both G' and η*, owing to the low absolute instrumental readings for a low viscosity material tested in this conditions, complex viscosity is still dependent on the deformation frequency whereas elastic modulus plots appear to be constant within the significant data scattering, thus suggesting a weak percolation network, accordingly with the poor dispersion of nanoflakes observed by electron microscopy.

Linear viscoelasticity in the molten state for pCBT/GRM nanocomposites was studied at 250°C due to the higher melting temperature of pCBT (~ 226°C) with respect to that of CBT (the highest melting peak is located at ~ 186°C). For all the pCBT/GRM nanocomposites, η* strongly depends on the deformation frequency, while the dependence of G' is weak, especially at low deformation frequencies, thus indicating the formation of a solid-like network for all the nanocomposites [40]. At low frequencies, the viscosity of pCBT + GNP exhibits a linear dependence with the frequency, with a η* value of ~ $10^3$ Pa·s (at 1 rad/s) which is one order of magnitude higher than that of pure pCBT. pCBT + RGO exhibits viscosity and G' values of ~ $6·10^3$ Pa·s and ~ $6·10^3$ Pa (at 1 rad/s), respectively, while the use of annealed RGO leads to η* ~ $6·10^4$ Pa·s and G' ~ $6·10^4$ Pa (at 1 rad/s), i.e. a factor of 10 increase for annealed nanoflakes. A much weaker increase of modulus and viscosity, in the range of 20%, was observed when comparing pCBT containing RGO-2 and RGO-2_1700. The increase of both G' and η* upon nanoflake annealing may be explained by an higher affinity of the polymer towards lower oxidized rGO, in agreement with data reported for PMMA/few layer graphene nanocomposites where higher impact on the viscoelastic properties were obtained with high C/O ratio nanoflakes [38]. However, such effect was not significant in CBT nanocomposites, in which negligible differences between G' and η* of nanocomposites containing pristine or annealed nanoparticles were observed.

To better estimate the dispersion degree of the different nanoflakes, fitting of viscosity data, for both CBT and pCBT nanocomposites, was carried out at the lower shear rates (ω = 0.1 ÷ 1 rad/s) with the aim of calculating the shear thinning exponent factor, n, of the equation

$$\eta = A \cdot \omega^n$$

where η is the viscosity, A is a sample specific pre-exponential factor and ω is the oscillation frequency of the rheometer; the value of n is supposed to be a semi-quantitative measure of the dispersion degree of the sample, as reported by Wagener and Reisinger [41]. Fitting results are plotted as straight lines in Figure 2a, and n values calculated for all the nanocomposites are reported in Table S2. Neat pCBT exhibits perfect Newtonian behavior at the employed shear rates, with a shear thinning exponent n = -0.05, in agreement with n calculated by Wagener and Reisinger for pure PBT [41]. The addition of GNP results in n = -0.74 and n = -0.68 for CBT and pCBT nanocomposites, respectively. When rGO are included in CBT and pCBT, higher n were calculated with values ranging between n = -0.87 for pCBT + RGO and n = -0.97 for CBT + RGO-2: this clearly indicates a higher dispersion degree of all rGO respect to GNP, in agreement with the above-discussed results.

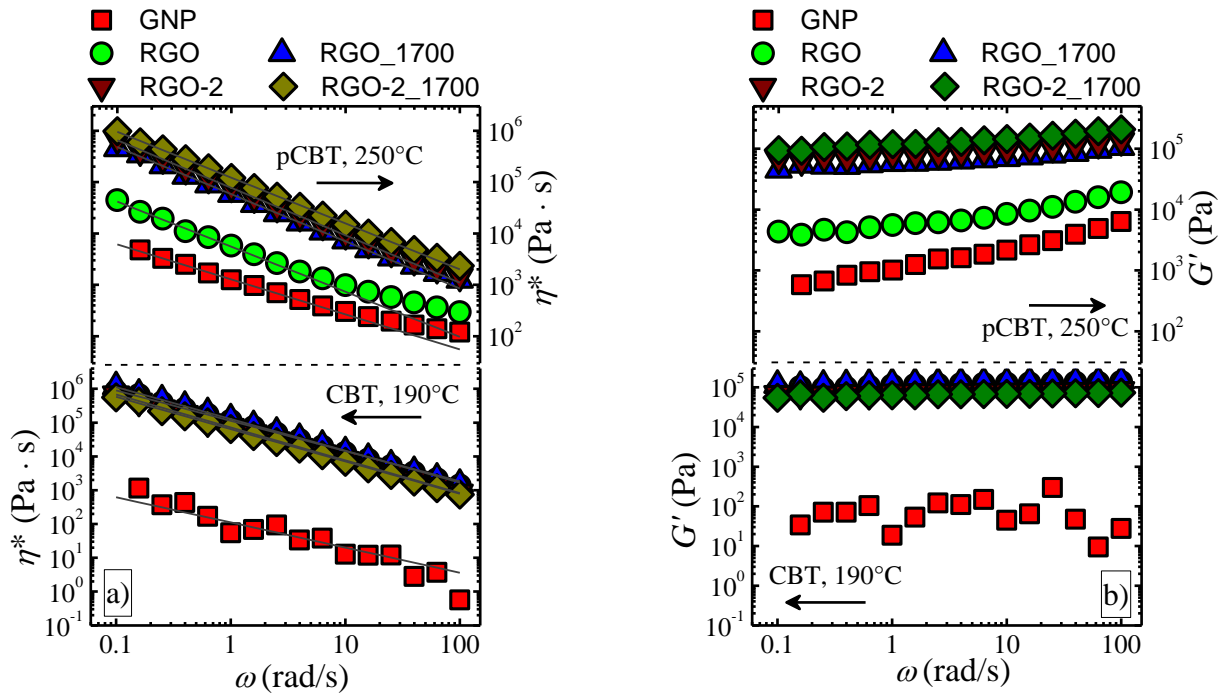

**Figure 2. Dynamic frequency sweep test at 190°C and 250°C for CBT and pCBT nanocomposites, respectively. (a) Complex viscosity and (b) G' as a function of the angular frequency. The straight lines in the panel (a) represents the fitting of the different curves to calculate the shear thinning exponent factor, n, as suggested by Wagener and Reisinger [41]**

Electron microscopy and linear viscoelasticity in the molten state showed the presence of a percolation network for all the nanocomposites; this, coupled with the intrinsic conductivity of graphene related materials is expected to result in electrically conductive composites. Indeed, while polymers are well known insulating materials with an extremely low electrical conductivity ($\sigma \sim 10^{-13}$ S/m for pure pCBT [26]) a sharp increase in the electrical conductivity has been typically observed upon dispersion of conductive nanoparticles at loading above the percolation threshold, which value depends primarily on particle aspect ratio and dispersion degree [26, 31].

Electrical conductivity results on CBT and pCBT nanocomposites, containing 5 wt.% of different graphene-related materials, are reported in Figure 3 and Table 1. Despite in literature no conductivity values were previously reported for CBT, an electrical conductivity of $\sim 10^{-13}$ S/m, *i.e.* equal to pCBT, can be taken as a realistic figure for CBT. Conductivity results for all prepared composites evidence for nanoflake percolation: pCBT nanocomposites range from $10^{-5}$ S/m for pCBT + GNP up to $1.3 \cdot 10^{-2}$ S/m for pCBT + RGO-2_1700. It is worth noting that electrical conductivity results are consistent with the rheological data above reported: indeed, rGO ($\sim 1.5 \cdot 10^{-4}$ and $\sim 1.0 \cdot 10^{-3}$ S/m when RGO and RGO-2, respectively, are used) are more efficient than GNP, pCBT + RGO-2 exhibits an higher electrical conductivity respect to pCBT + RGO, while the electrical conductivities of pCBT + RGO_1700 ($\sim 9.0 \cdot 10^{-4}$ S/m) and pCBT + RGO-2 display similar values; however, the use of annealed rGO, compared to their pristine counterparts, leads to electrical conductivity values of about one order of magnitude higher, reflecting both the higher dispersion degree and the lower defectiveness of thermally treated nanoflakes. Comparing nanocomposites based on CBT and pCBT, similar trends on the electrical conductivity values were clearly obtained for the different GRM: CBT + GNP is the nanocomposite with the lowest electrical conductivity ($\sim 4.9 \cdot 10^{-3}$ S/m) while the use of annealed rGO leads to higher values respect to their pristine counterparts ($\sim 1.7 \cdot 10^{-2}$ and $\sim 4.7 \cdot 10^{-2}$ S/m when RGO and RGO_1700, respectively, were added to CBT, while for RGO-2 values of $\sim 5.4 \cdot 10^{-3}$ and $\sim 1.5 \cdot 10^{-1}$ S/m were measured for CBT + RGO-2 and CBT + RGO-2_1700, respectively). However, from the direct comparison of conductivities of CBT/GRM *vs.* pCBT/GRM, unpolymerized nanocomposites are systematically more electrically conductive than their pCBT/GRM counterparts, due to a dramatic reduction of nanoflake aspect ratio [42] upon longer melt blending time and viscosity increase occurring during polymerization, as demonstrated by the electron image analysis on RGO_1700 nanoflakes extracted from their relevant CBT and pCBT nanocomposites (Figure S4).

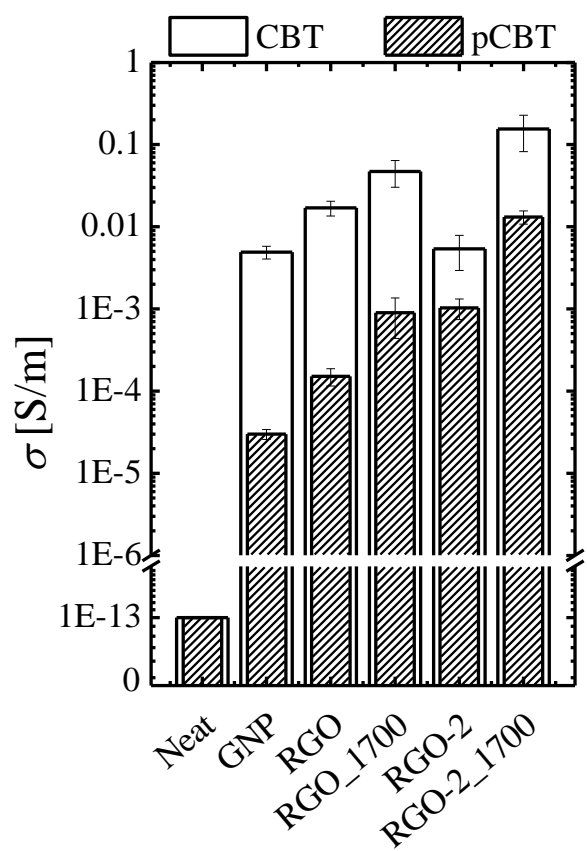

**Figure 3. Electrical conductivity vs. nanoparticle and matrix type. The filler content is set constant at 5 wt.%. The value here reported for pure pCBT was taken from ref. [26] while value for pure CBT was supposed to be equal to that of pure pCBT.**

Polymers are well known thermally insulating materials, with typical thermal conductivity values in the range 0.1 ÷ 0.4 W/(m·K) [1] and the addition of thermally conductive nanoparticles is known to have a positive effect for the improvement of this property. Thermal conductivity ($\lambda$) results for CBT and pCBT nanocomposites containing 5 wt.% of different graphene related materials are reported in Table 1 while the relative increase in conductivity ($\lambda_{nanocomposite}/\lambda_{matrix}$) are shown in Figure 4. Pure CBT and pCBT have thermal conductivities of ~ 0.22 and ~ 0.24 W/(m·K), respectively, which is consistent with typical values of semi-crystalline polymers [1]. The addition of 5 wt.% of GNP, RGO and RGO-2 has a limited effect on thermal conductivity of either CBT and pCBT with values in the range of ~ 0.4 ÷ 0.5 W/(m·K), *i.e.* about twice those of pure oligomer/polymer. On the other hand, the use of annealed nanoparticles dramatically increases the thermal conductivity of nanocomposites up to ~ 1.48 ± 0.02 W/(m·K) for CBT + RGO_1700, which is about 7 times the value measured for pure CBT: such a high enhancement in the thermal conductivity of nanocomposites containing annealed rGO was previously correlated to the lower defectiveness of annealed rGO, combined with their high aspect ratio [18]. The novel aspect from the present paper is that thermal conductivity values obtained for CBT/rGO_1700 nanocomposites are ~ 20 ÷ 65 % higher than for the correspondent formulations based on pCBT. This is likely related to the higher aspect ratio of nanoflakes in CBT nanocomposites, as the effect of aspect ratio reduction was previously recognized to be detrimental in polymer nanocomposites [2]. In fact, the progressively increasing applied shear during melt mixing, owed to the viscosity increase during ring-opening polymerization of CBT into pCBT, is responsible for the reduction of nanoflake aggregate size, as demonstrated in the case of RGO_1700 nanoflakes extracted from their relevant CBT and pCBT nanocomposites (Figure S4).

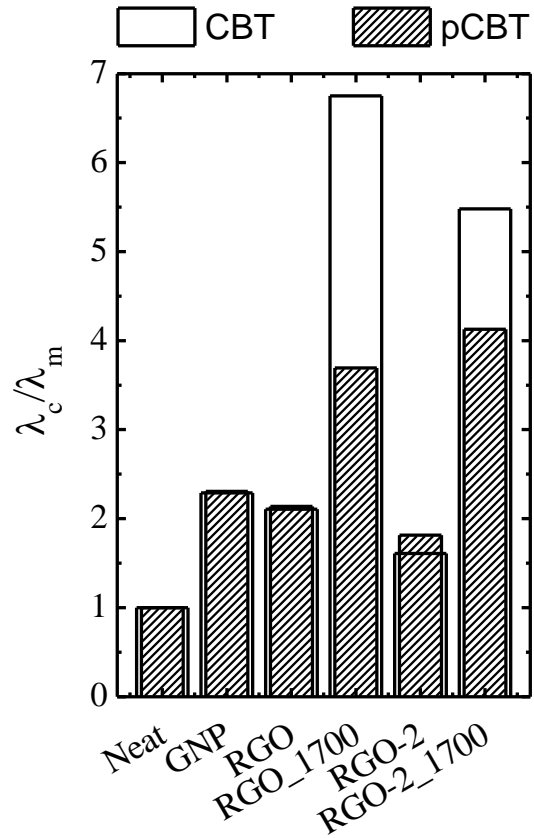

**Figure 4. Normalized thermal conductivity vs. nanoparticle and matrix type. The filler content is constant at 5 wt.%.**

**Table 1. Electrical and thermal conductivity data for CBT and pCBT nanocomposites**

| Nanoparticle Type | CBT | pCBT |
|---|---|---|
| | Electrical conductivity [S/m] | |
| | Thermal Conductivity [W/(m·K)] | |
| | $\sigma \sim 1.0$ E-13 | $\sigma \sim 1.0$ E-13 |
| | $\lambda = 0.219 \pm 0.005$ | $\lambda = 0.241 \pm 0.001$ |
| GNP | $\sigma = (4.9 \pm 0.1)$ E-3 | $\sigma = (3.0 \pm 0.4)$ E-5 |

|  | λ = 0.501 ± 0.002 | λ = 0.556 ± 0.001 |
|---|---|---|
| RGO | σ = (1.7 ± 0.3) E-2 | σ = (1.5 ± 0.4) E-4 |
|  | λ = 0.461 ± 0.001 | λ = 0.515 ± 0.004 |
| RGO_1700 | σ = (4.7 ± 1.7) E-2 | σ = (9.0 ± 4.6) E-4 |
|  | λ = 1.478 ± 0.020 | λ = 0.890 ± 0.009 |
| RGO-2 | σ = (5.4 ± 2.5) E-3 | σ = (1.0 ± 0.3) E-3 |
|  | λ = 0.352 ± 0.001 | λ = 0.437 ± 0.003 |
| RGO-2_1700 | σ = (1.5 ± 0.7) E-1 | σ = (1.3 ± 0.2) E-2 |
|  | λ = 1.200 ± 0.007 | λ = 0.995 ± 0.001 |

The use of high aspect ratio and low defective rGO turned out to be the best combination for the preparation of highly thermally and electrically conductive polymer nanocomposites, both in CBT and pCBT. Furthermore, the infiltration, and the consequent disaggregation, of rGO aggregates turned out to be mandatory for the preparation of highly conductive nanocomposites. Indeed, pCBT + RGO nanocomposites prepared by direct melt blending pCBT and RGO, *i.e.* without the pre-dispersion of RGO in CBT, exhibited poor electrical and thermal conductivity values ($\sigma \approx 2 \cdot 10^{-5}$ S/m and $\lambda \approx 0.37$ W/m·K). On the other hand, nanocomposites prepared by solvent assisted pre-mixing pCBT and RGO led to electrical conductivity ($\sigma \approx 3 \cdot 10^{-4}$ S/m), while thermal conductivity ($\lambda \approx 0.45$ W/m·K) close to values reported above for pCBT +RGO (detailed comments in SI).

Finally, it is worth mentioning that the low molecular weight of CBT oligomers (and the associated low mechanical properties) prevent the use of unpolymerized CBT nanocomposites in most practical applications. However, CBT nanocomposites may find applications as intermediates in polymer processing, *e.g.* as masterbatches in the preparation of nanocomposites with low nanoparticles content.

## 4. Conclusions

The present paper is focused on morphology and conductivity properties evolution upon ring-opening polymerization during extrusion for the production of poly butylene terephthalate nanocomposites containing graphene-related materials, including GNP, rGO and thermally annealed rGO. Despite unpolymerized CBT nanocomposites cannot be regarded as materials for real application, owed to their low mechanical properties and low melting temperature range, this work was aimed at elucidating how nanocomposite properties change upon the ring-opening polymerization. This can help to develop and optimize reactive extrusion processing of pCBT nanocomposites in order to improve the desired properties.

The extremely low viscosity of CBT during compounding did not allow to apply sufficiently high shear forces during melt mixing, resulting in poor dispersion and distribution degree of nanoparticles. However, the low viscosity allowed the infiltration of oligomers into the accordion-like structure of rGO aggregates. Viscosity increase during polymerization of CBT into pCBT, resulting in a rise of the applied shear during melt mixing, led to a higher dispersion and distribution degree of rGO nanoflakes, as well as their lateral size reduction. Linear viscoelasticity in the molten state showed the presence of a percolation network for all the nanocomposites. However, clear differences are observed between nanocomposites containing rGO and GNP, in terms of higher density of the percolation network obtained for rGO nanoflakes.

Electrical conductivity results for CBT and pCBT nanocomposites were consistent with the rheological data, the conductivity with rGO being typically significantly higher compared to GNP. Furthermore, the conductivity with annealed rGO being greater compared to pristine rGO for both CBT and pCBT nanocomposites (the highest values $\sigma \approx 0.2$ S/m and $\sigma \approx 0.01$ S/m for CBT and pCBT nanocomposites, respectively, containing annealed RGO-2) evidencing the importance of the exploitation of nanoflakes with low defectiveness. Electrical conductivities of CBT based materials are systematically higher (up to two order of magnitude for GNP, RGO and RGO_1700) respect to pCBT nanocomposites. This is related to the reduction of aspect ratio of nanoflakes upon polymerization, owing to the longer processing time and the rising applied shear as a consequence of viscosity increase during polymerization of CBT into pCBT.

The thermal conductivity of nanocomposites was strongly affected by the quality of nanoflakes. The use of as obtained/s received GNP and rGO resulted in nanocomposites with limited thermal conductivity improvements (~ 2-fold increase) independently on the matrix, while the higher values (ranging between 3.5 and 7-fold increase) were obtained with annealed rGO, further confirming the importance of exploitation of high quality graphene-related materials. Comparison between CBT and pCBT nanocomposites, containing annealed rGO, showed better thermal conductivities for CBT nanocomposites (the highest values ~ 1.5 W/(m·K) and ~ 1.0 W/(m·K) for CBT and pCBT nanocomposites, respectively), which is consistent with electrical conductivities and related to the mentioned reduction in aspect ratio of nanoflakes upon polymerization. These results provide for the first time, to the best of author's knowledge, experimental evidences of the reduction of aspect ratio of graphene related materials during reactive extrusion and its effect on the thermal conductivity of the relevant polymer nanocomposites.

**Authors contributions**

A. Fina conceived the experiments and coordinated the project, S. Colonna carried out the entire preparation of nanocomposites and most of the characterizations reported in this paper. O. Monticelli contributed to ring opening polymerization design and characterization. J. Gomez synthetized GNP and RGO. G. Saracco contributed to the discussion of the results. Manuscript was written by S. Colonna and A. Fina.

**Acknowledgements**

The research leading to these results has received funding from the European Union Seventh Framework Programme under grant agreement n°604391 Graphene Flagship. This work has received funding from the European Research Council (ERC) under the European Union's Horizon 2020 research and innovation programme  grant agreement 639495 — INTHERM — ERC-2014-STG. Funding from Graphene@PoliTo initiative of the Politecnico di Torino  is also acknowledged.

The authors gratefully acknowledge Dr. Mauro Raimondo for FESEM observations and Fausto Franchini for electrical conductivity measurements. Prof. Matteo Pavese is also gratefully acknowledged for high temperature annealing treatments.

# Morphology and properties evolution upon ring-opening polymerization during extrusion of cyclic butylene terephthalate and graphene-related-materials into thermally conductive nanocomposites

S. Colonna[a], O. Monticelli[b], J. Gomez[c], G. Saracco[d], A. Fina[a,*]

[a]Dipartimento di Scienza Applicata e Tecnologia, Politecnico di Torino, 15121 Alessandria, Italy
[b]Dipartimento di Chimica e Chimica Industriale, Università di Genova, Via Dodecaneso, 31, 16146 Genova, Italy
[c]AVANZARE Innovacion Tecnologica S.L., 26370 Navarrete (La Rioja), Spain
[d]Istituto Italiano di Tecnologia, Centre for Sustainable Futures CSF@PoliTo, 10129 Torino, Italy
*Corresponding author: alberto.fina@polito.it

**Differential scanning calorimetry**

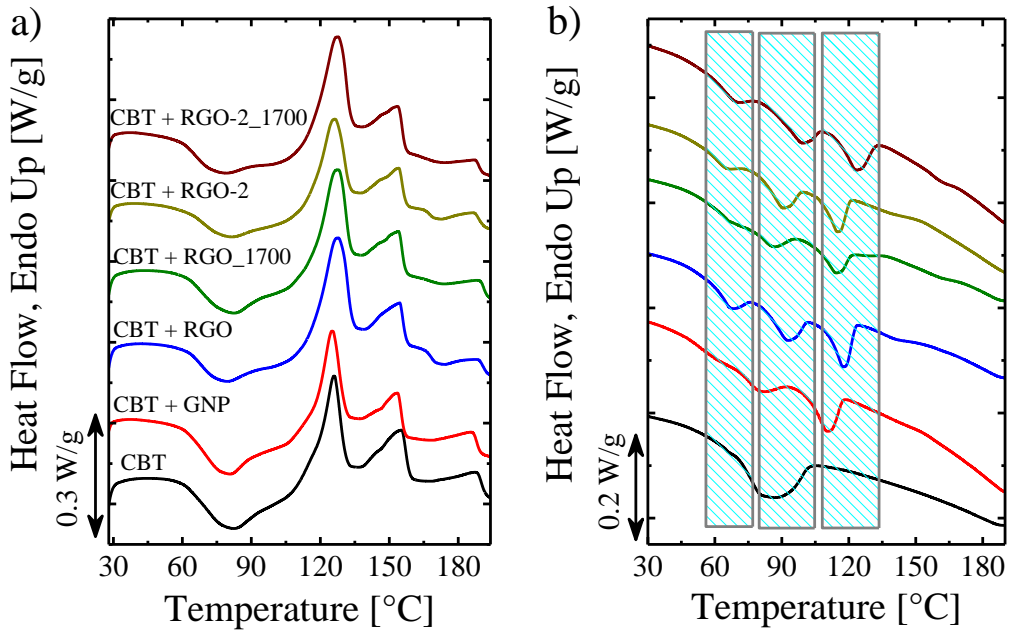

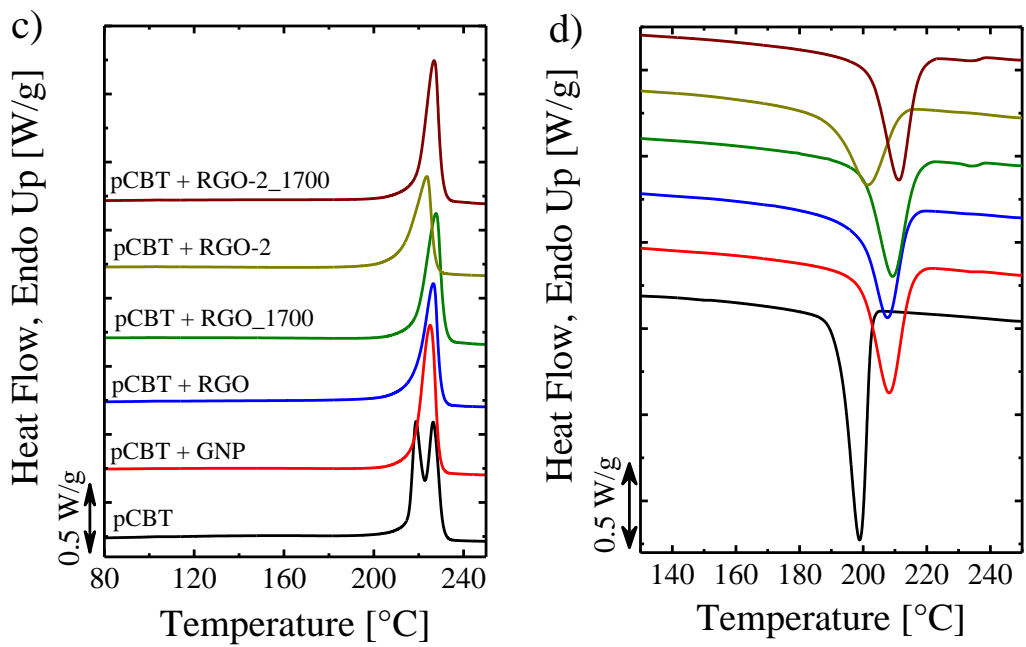

Figure S1. DSC results on CBT and pCBT nanocomposites: a) 2^nd heating and b) cooling for CBT nanocomposites; c) 2^nd heating and b) cooling for pCBT nanocomposites

**Comparison of processing conditions**

Two additional pCBT + 5% RGO nanocomposite were prepared, based on two different preparation protocols, to compare the effect of these preparation methods, and of the "standard" *in-situ* polymerization, on the properties of pCBT + 5% RGO.

Protocol A was designed to compare the effect of viscosity during the infiltration into the accordion-like structure of RGO, without changing any parameter other than the molecular weight, i.e. pCBT vs CBT:

- CBT was extruded for 10 minutes at 250°C and 100 rpm in presence of tin catalyst, thus obtaining pCBT;
- pCBT was completely dissolved under stirring pellets in $CHCl_3$/HFIP (1/1 v/v) mixture; then, 5 wt.% RGO was added to the solution and the solvent evaporated.
- The dried mixture was pulverized, added into the extruded and mixed for 10 minutes at 250°C and 100 rpm and the obtained nanocomposite named PBT + RGO.

.

Protocol B was designed as a simple melt blending reference:

- CBT was extruded for 10 minutes at 250°C and 100 rpm, in inert atmosphere, in presence of tin catalyst, thus obtaining pCBT;
- Then screw rotation speed was reduced down to 30 rpm and RGO (5 wt.% with respect to the pCBT content) was added directly into the extruder;
- Extrusion was carried out for further 10 minutes at 250°C and 100 rpm. The material here obtained is named (PBT + RGO)_sf, where sf indicates solvent free

Scanning electron microscopy at low magnitude revealed the presence of aggregates in the range of some tens of microns when the nanocomposites was prepared "solvent-free" (Figure S3e) whereas no aggregates at this magnification were observed for pCBT + RGO (Figure S3a) and PBT + RGO (Figure S3c). High magnification micrographs showed regular distribution of nanoflakes (highlighted by white arrows) for both nanocomposites prepared via solvent-assisted pre-mixing (Figure S3b,d), whereas in (PBT + RGO)_sf (Figure S3f) a very limited number of nanoflakes is visible, likely due to the poor disaggregation of RGO aggregates. These results suggest that direct melt blending is not able to disaggregate RGO expanded structure, leading to a limited dispersion and distribution of nanoflakes. On the other hand pre-infiltration of oligomers/polymer chains into the galleries of expanded rGO structure was demonstrated to result in a significantly improved dispersion and distribution of rGO nanoflakes.

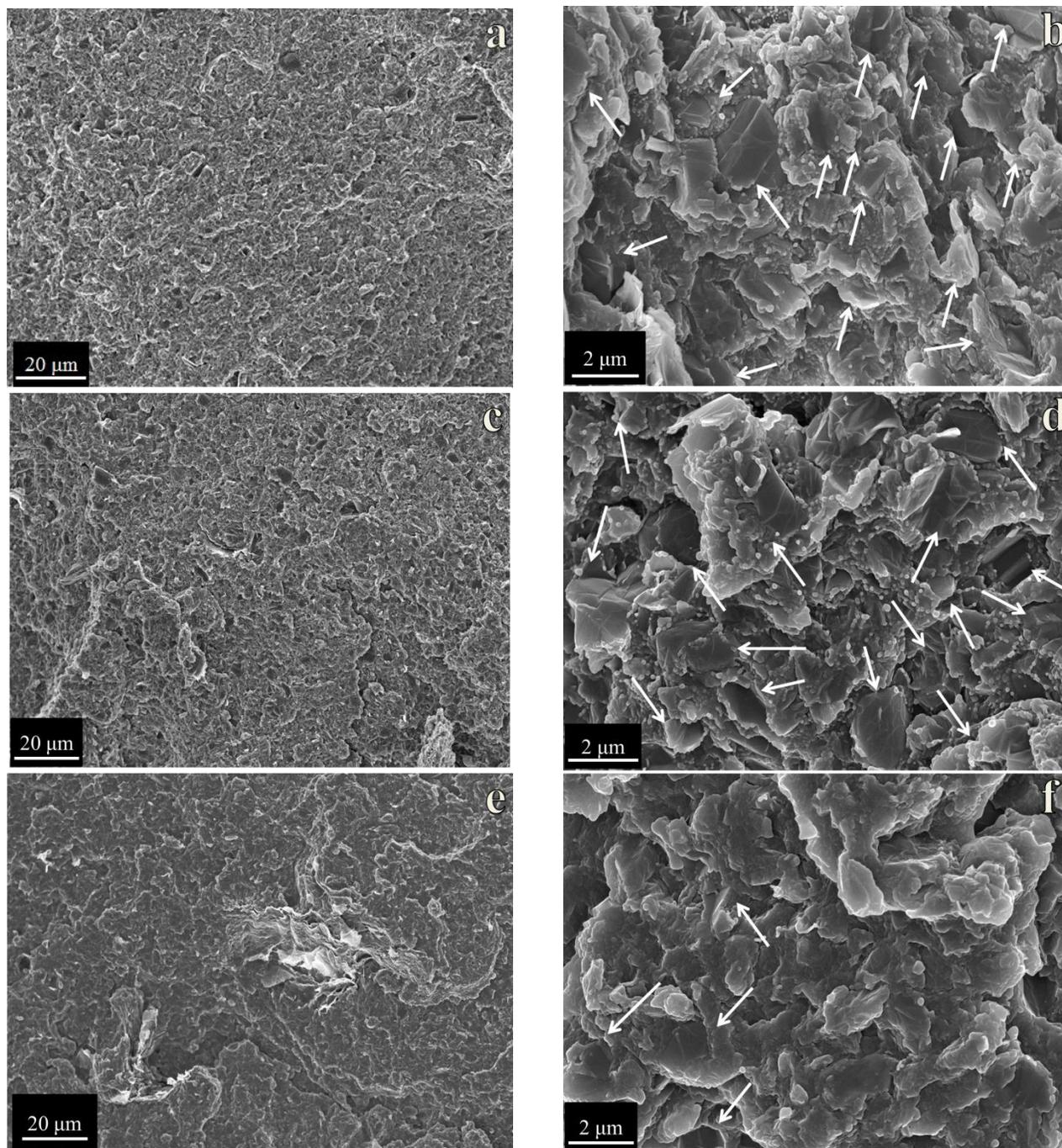

Figure S2. Electron microscopy low and high magnification for (a,b) pCBT + RGO, (c,d) PBT + RGO and (e,f) (PBT+RGO)_sf

Furthermore, thermal and electrical conductivity of the nanocomposites were studied (Table S1). As expected from the poor distribution and dispersion, PBT + RGO_sf resulted in low electrical ($\sigma \approx 2 \cdot 10^{-5}$ S/m) and thermal conductivity ($\lambda \approx 0.37$ W/mK). Comparing pCBT + RGO with PBT + RGO, similar electrical conductivity were obtained (in the order of $10^{-4}$ S/m), whereas thermal conductivity was found to be slightly higher for pCBT + RGO (0.51 W/mK) than for PBT + RGO (0.45 W/mK).

SEM observation and thermal/electrical conductivities are therefore consistent in demonstrating the need to infiltrate oligomers or polymers, via solvent processing, to allow a proper dispersion of nanoflakes, which in turn results in higher conduction performance. Slight differences apply between the infiltration of CBT oligomers or pCBT.

Table S1. Electrical and thermal conductivity for pCBT/RGO nanocomposites prepared following different mixing procedures.

|  | $\lambda$ [W/(m·K)] | $\sigma$ [S/m] |
|---|---|---|
| pCBT + RGO | 0.515 ± 0.004 | (1.5 ± 0.4) E-4 |
| PBT + RGO | 0.454 ± 0.004 | (3.2 ± 1.1) E-4 |
| (PBT + RGO)_sf | 0.367 ± 0.001 | (2.4 ± 0.2) E-5 |

**Electron microscopy for CBT vs pCBT nanocomposites**

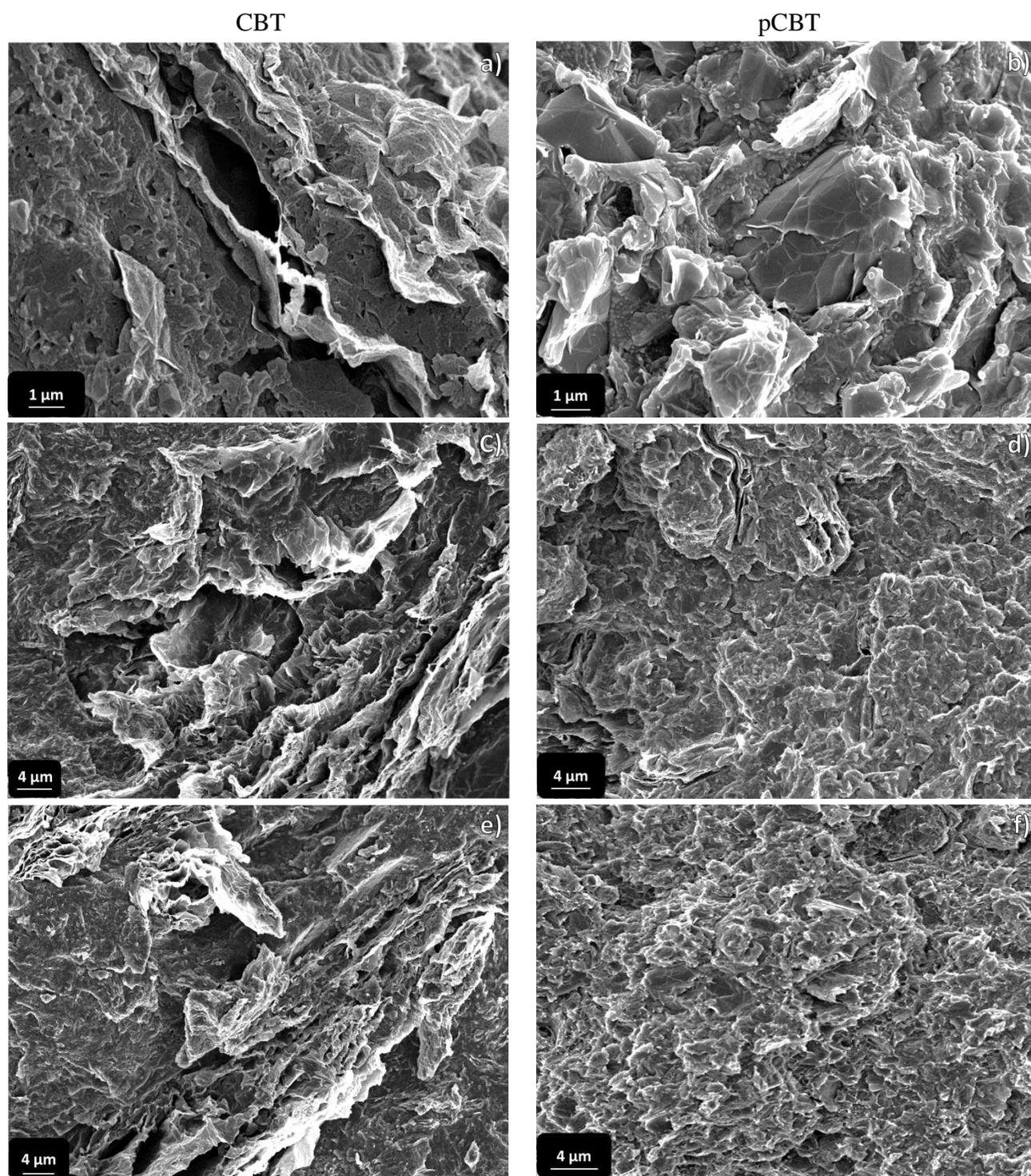

Figure S3. FESEM micrographs of a) CBT + RGO_1700, b) pCBT + RGO_1700, c) CBT + RGO-2, d) pCBT + RGO-2, e) CBT + RGO-2_1700 and f) pCBT + rGO-2_1700.

## Rheology

Table S2. Rheological results for CBT and pCBT nanocomposites

|  | $\eta^*$ [Pa · s] @ $\omega$ = 1 rad/s<br>G' [Pa] @ $\omega$ = 1 rad/s<br>n | |
|---|---|---|
|  | **CBT** | **pCBT** |
| **GNP** | $\eta^*$ = 5.4 E1<br>G' = 1.9 E1<br>n = -0.74 | $\eta^*$ = 1.25 E3<br>G' = 1.00 E3<br>n = -0.68 |
| **RGO** | $\eta^*$ = 1.13 E5<br>G' = 1.11 E5<br>n = -0.90 | $\eta^*$ = 5.80 E3<br>G' = 5.50 E3<br>n = -0.87 |
| **RGO_1700** | $\eta^*$ = 1.30 E5<br>G' = 1.28 E5<br>n = -0.94 | $\eta^*$ = 5.79 E4<br>G' = 5.74 E4<br>n = -0.91 |
| **RGO-2** | $\eta^*$ = 7.58 E4<br>G' = 7.53 E4<br>n = -0.97 | $\eta^*$ = 7.94 E4<br>G' = 7.89 E4<br>n = -0.92 |
| **RGO-2_1700** | $\eta^*$ = 6.42 E4<br>G' = 6.31 E4<br>n = -0.95 | $\eta^*$ = 1.20 E5<br>G' = 1.19 E5<br>n = -0.89 |

## Particle size analysis

Particle size analysis of nanoflakes extracted form nanocomposites was performed analyzing RGO_1700 flakes, deposited on Si wafer, by means of scanning electron microscope. Granulated CBT + RGO_1700 and pCBT + RGO_1700 were dissolved in tetrahydrofuran (THF) and chloroform/hexafluoroisopropanol ($CHCl_3$/HFIP) mixture (90/10 v/v), respectively, for 2 hours under stirring; then, the suspension was vacuum filtered (0.45 μm pore size) to separate solubilized polymer from nanoflakes. Finally, after drying the filter for two hours in oven at 80°C, nanoflakes were separated from the filter and collected in a glass vial.

Later ~ 0.1 mg of RGO_1700, for nanoflakes obtained from both CBT and pCBT nanocomposites, were dissolved in ~ 10 ml $CHCl_3$/HFIP mixture and sonicated in bath for 30 minutes to suspend nanoflakes; then, the suspensions were drop-casted on a silicon wafer and the solvent evaporated under a chemical hood. Deposited nanoflakes were observed without any further preparation. Particle size analysis was performed by means of image analysis software evaluating the projected area on more than 50 nanoflakes. Results for RGO_1700 obtained from CBT and pCBT are reported in Figure S3.

Distribution of projected area of RGO_1700 nanoflakes, extracted from CBT nanocomposites, display an approx. lorentzian curve with a maximum at 54 μm and 50% of nanoparticles showing an area below 16 µm², while for those extracted from pCBT nanocomposites a different distribution of projected area was observed, with a dramatic increase of the fraction of nanoflakes with small area, leading to 50% of nanoparticles showing an area smaller than 4 µm². It is worth noting that in both cases, the projected area is typically related to rGO aggregates, even if the smaller area observed. Smaller and thinner individual nanoflakes are difficult to be detected in these conditions and may be underestimated in this analysis. Nonetheless, significant differences are visible between size distribution in CBT and pCBT, suggesting higher disaggregation of nanoflake aggregates in pCBT, owing to its higher viscosity.

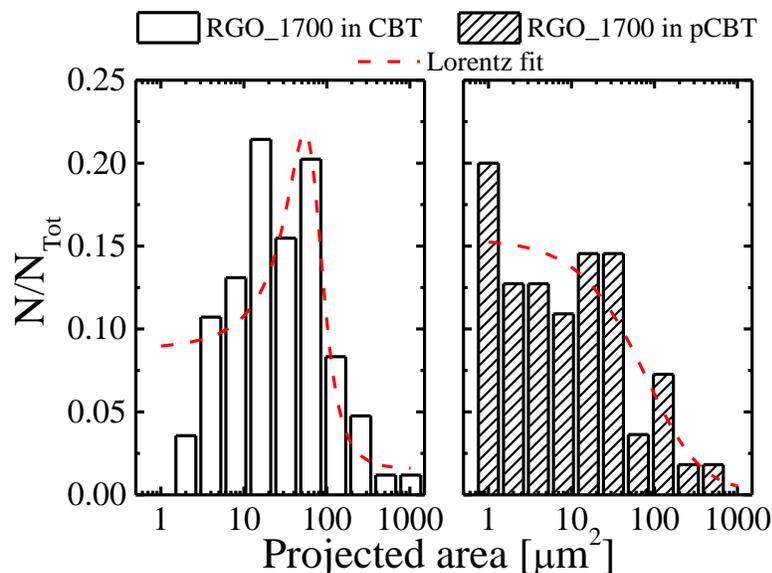

Figure S4. RGO_1700 projected area distribution before and after ring-opening polymerization of CBT into pCBT